# Visualizing modified spin-wave wavefronts near magnetic defects and domains using nitrogen-vacancy centers

Wenxin Cheng[1,2,†], Chang Liu[1,†,*], Dekun Shen[3,4,†], Jiaxin Li[1,2,†], Shangyuan Wang[3,4], Hongyu Wang[1], Miming Cai[3,4], Jihao Xia[1], Peng Chen[1], Caihua Wan[1], Ka Shen[3,4], Xiufeng Han[1], Yuelin Zhang[3,4,*], Jinxing Zhang[3,4], Yangmu Li[1,2,*]

[1]*Beijing National Laboratory for Condensed Matter Physics, Institute of Physics, Chinese Academy of Sciences, Beijing 100190, China*
[2]*School of Physical Sciences, University of Chinese Academy of Sciences, Beijing 100049, China*
[3]*School of Physics and Astronomy, Beijing Normal University, Beijing 100875, China*
[4]*Key Laboratory of Multiscale Spin Physics, Ministry of Education, Beijing 100875, China*

†: These authors contribute equally to this work
*: Correspondence authors:
yangmuli@iphy.ac.cn; yuelin.zhang@bnu.edu.cn; liuchang@iphy.ac.cn

**Direct, real-space imaging of spin-wave propagation and wavefronts in magnetic materials is crucial for advancing both fundamental understanding of spin dynamics and the development of functional devices. This, however, remains a significant challenge, especially in materials with complex magnetic characteristics at the nanoscale. Here, we employ scanning nitrogen-vacancy center spectroscopy to achieve visualization of spin waves in two archetypical magnetic films: yttrium-iron-garnet and lanthanum strontium manganese oxide. We reveal a wavelength-dependent spin-wave filtering effect near point-like magnetic scatterers and a modified spin wavefront in antiferromagnetically coupled stripe domains. The spin-wave characteristics are explained using micromagnetic simulations and analytical calculations. These findings point to possible fine control of spin-wave propagation near complex magnetic structures and extend the scope of spin-wave imaging based on nitrogen-vacancy centers beyond uniform magnets.**

The design and control of spin waves in magnetic structures are a central challenge in spintronics and magnonics, critical for developing low-power, wave-based computation and information processing technologies [1-3]. Spin waves (magnons), the collective excitations of magnetic order, offer a solution to the Joule heating that is detrimental to charge-based electronic devices [4-7]. This advantage is particularly pronounced in low-damping magnetic materials, where spin waves propagate with minimal energy loss [8,9], and it offers a novel pathway for realizing nanoscale spin-wave logic and non-volatile memory [1-3,10,11]. Detailed investigations of spin-wave behaviors near complex magnetic structures represent a crucial step toward these functional devices.

Despite the application potential of engineered spin waves, a better understanding of their interaction with complex magnetic structures is needed for fine control of spin-wave behaviors and logic operations [12-14]. Near magnetic scatterers or at domain boundaries, a spin wave can be scattered, reflected, and phase-shifted. Its wavefront and transmission probability become frequency-dependent, and it can experience substantial energy and momentum changes. These effects originate from the local variations in the effective magnetic field, which modifies the spin-wave dynamics governed by the Landau-Lifshitz-Gilbert (LLG) equation [15,16]. Previously, a vast body of knowledge about spin waves has been obtained by experimental tools including Brillouin light scattering [17], scanning transmission X-ray microscopy [18-20], and time-resolved magneto-optical Kerr effect [21], which have provided detailed characterizations of key physical properties of spin waves. Fundamentally, these physical properties of spin waves (excitations of the spin lattice) are determined by the local spin patterns in the magnetic ground state.

In this work, using scanning nitrogen-vacancy (NV) center spectroscopy, we directly visualize spin waves near ferromagnetic scattering centers in yttrium-iron-garnet and within antiferromagnetically coupled stripe domains in lanthanum strontium manganese oxide. Using resonance between spin quantum levels, an NV center can provide information complementary to existing techniques by investigating the phase propagation, handedness, and dynamics of spin waves while simultaneously measuring magnetic textures. Near the scattering centers, we observe an interference pattern analogous to classical wave scattering. Within the stripe domains, we discover a distinct modification of spin waves. The motion of ferromagnetic scattering centers and antiferromagnetically coupled stripe domains is negligible during the experiment, and thus both can be viewed as quasi-static local variations of the magnetic environment.

Yttrium-iron-garnet ($Y_3Fe_5O_{12}$, YIG) [9] and lanthanum strontium manganese oxide ($La_{1-x}Sr_xMnO_3$, LSMO) [22,23] films were synthesized for spin-wave measurements. YIG (100-nm-thick) films were epitaxially grown on $Gd_3Ga_5O_{12}$ substrates via magnetron sputtering, with a [111] out-of-plane orientation and in-plane edges along the [110] direction [24-26]. YIG films exhibit a Curie temperature of approximately 560 K [9]. At our measurement temperature, our YIG film is in a macroscopically uniform, single-domain state, free of domain walls. LSMO films with $x = 0.33$ (160 nm thickness) were grown on $NdGaO_3$ substrates with the orientation of [110] by pulsed laser deposition technique [27,28]. LSMO is a half-metallic ferromagnet with a Curie temperature of approximately 370 K [23,28], which also exhibits very low Gilbert damping [29]. Due to the Dzyaloshinskii-Moriya interaction, complex magnetic textures can be stabilized in LSMO [27,30]. By initializing with an in-plane field and subsequently removing it, spin stripes of upward and downward domains can form, with a domain periodicity associated to the film thickness [28]. These domains function as two antiferromagnetically coupled regions that host spin-wave modes both within the stripes and at the boundaries [28,31].

Our experimental setup is illustrated in Fig. 1(a), where a [100]-cut diamond tip with an optical waveguide array containing NV centers [32] is mounted on a tuning fork that scans over the film surface [33,34]. The NV centers are prepared to have a spin quantum number of $S = 1$, which results in a ground state with $m_s = 0, \pm 1$ sublevels [35-37]. By optically pumping the NV spins and letting them decay into the ground state, spin-projection-related fluorescence is emitted. A straight microstrip gold antenna is deposited onto the magnetic films to generate spin-wave excitations (supplementary Fig. S1 [38]). The excited spin-wave vector depends on the width of the antenna, and the direction of the wave vector depends on the antenna geometry, which are designed to separate multiple spin modes. A dual-way confocal objective lens focuses the pumping (532 nm) laser, visible light, and emitted fluorescence onto and from the optical waveguide array [33,34]. In our experiments, an external field is applied along the NV-axis to control the spin wave excitation and lift the degeneracy of NV's $m_s = \pm 1$ sublevels, and all measurements in this work are obtained by resonating between NV's $m_s = 0$ and -1 sublevels. Microwaves are applied to both resonate NV's spins and to serve as a reference electromagnetic field for the spin-wave imaging [46-53]. Since the reference field in air has a much longer periodicity compared to spin waves in magnetic films (i.e., mm vs. μm), it can be considered as an electromagnetic field that is spatially uniform in phase but time-dependent. Spin waves with the same frequency of the reference field are stabilized for the imaging, filtering out components with other frequencies. This method does not require reconstructing magnetization from stray fields. No microfabrication of the sample is required prior to NV measurements.

For a propagating spin wave with a wavevector $\boldsymbol{k}$, frequency $\omega$, positive group velocity, and a precession magnetization $\boldsymbol{M}$ in a uniform ferromagnetic material with a saturation magnetization $M_s$, its position and time-dependent magnetization can be written as [48,49]:

$$\boldsymbol{M}(\boldsymbol{\rho}, t) = m e^{i(-\boldsymbol{k}\cdot x + \omega t)}(\hat{e}_x - \hat{e}_z i\eta) + \hat{e}_y M_s \qquad (1)$$

where $\boldsymbol{\rho} = x\hat{e}_x + y\hat{e}_y$ labels the in-plane 2D coordinate vector. $m$ is the amplitude of the dynamic magnetization component. Unit vectors are $\hat{e}_x$, $\hat{e}_y$, and $\hat{e}_z$. In this work, the spin wave propagates along $x$-axis. Gold antenna is along $y$-axis and the normal direction of magnetic films is along $z$-axis. For our YIG films, the equilibrium magnetization is approximately along the external field direction $y$ [48,49]. $t$ labels the time. $\eta$ is a polarization factor ranging between 0 to 1. The generalized Rabi frequency $\Omega$ is given by[48,49]:

$$\Omega(\boldsymbol{\rho}) = \sqrt{\Omega_R^2(\boldsymbol{\rho}) + \delta^2}$$

$$\Omega_R(\boldsymbol{\rho}) = \frac{\gamma}{2\sqrt{2}\hbar}\left\{\begin{matrix} B_{\mathrm{ref}}^2 + B_{\mathrm{SW}}^2\left[\left(\cos\theta_{NV}\sin\phi_k\right)^2 + \left(\cos\phi_k + \sin\theta_{NV}\right)^2\right] \\ -2B_{\mathrm{ref}}B_{\mathrm{SW}}\left[\cos\boldsymbol{k}\cdot\boldsymbol{\rho}\left(\cos\phi_k + \sin\theta_{NV}\right) - \sin\boldsymbol{k}\cdot\boldsymbol{\rho}\cos\theta_{NV}\sin\phi_k\right]\end{matrix}\right\}^{1/2} \qquad (2)$$

where $\delta$ is detuning frequency from NV's sublevel resonance. $\gamma$ is the gyromagnetic ratio. $B_{\mathrm{ref}}$ and $B_{\mathrm{SW}} = \frac{\mu_0}{2} e^{-|\boldsymbol{k}|d}(1 - e^{-|k|L}) m\left(\cos\phi_k + \eta\right)$ are GHz reference and spin wave field amplitudes, respectively. $\mu_0$ is magnetic permeability in vacuum. $d$ is the stand-off distance of the NV center from the magnetic films. $L$ is film thickness. $\theta_{NV}$ is the tilting angle of the NV-axis from the normal direction of the magnetic films. $\phi_k$ is the in-plane angle of spin-wave vector $\boldsymbol{k}$. The spin waves are measured via $\Omega$ between NV centers' spin states, which can be independently obtained from the static local field (energy shift of the spin states). When $B_{\text{ref}}$ and $B_{\text{SW}}$ are comparable to each other, the variations in $\Omega$ are most prominent, leading to a better sensitivity achieved by selecting the height $d$ during measurements. The photoluminescence (PL) contrast of the emitted fluorescence can be written as:

$$\text{PL contrast} = C\left(\frac{\delta^2}{\Omega_{\text{R}}^2(\boldsymbol{\rho})+\delta^2} - 1\right) \quad (3)$$

where $C$ is relevant to the PL contrast between coherent and fully occupied $m_{\text{s}} = 0, \pm 1$ sublevels. The contrast varies between 0 (not at NV resonance) and roughly $-C$. For a typical experiment $C \approx 0.15$, as PL of $m_{\text{s}} = \pm 1$ sublevels is approximately 30 percent weaker than that of the $m_{\text{s}} = 0$ sublevel. We note that in our experiment, we used a continuous laser pumping, which results in a dynamic balance of spin distribution between the NV's sublevels and it leads to a laser and microwave power-dependent $C$, an effect neglected in Eq. (3) [54]. For a spin wave with a simple phase structure (e.g., a plane wave), its phase can be solved from Eq. (2). For interference patterns or near magnetic domains, inhomogeneous background affects the solution and leads to a multiple-value problem. In this case, one may determine the spin-wave phase using the Riesz transform (two-dimensional Hilbert transform). Riesz transform enables the extraction of a local phase by constructing a monogenic signal from PL contrast. Detailed derivation is included in supplementary materials [38].

For typical uniformly magnetized materials, the local field variations are very small compared to the applied external magnetic field, and the PL contrast measured at a single frequency (resonance frequency corresponding to the external field) can be used to extract the local phase. However, for materials with magnetic domains, local field variations cause a change in the NV's resonance frequency, resulting in a location-dependent $\omega_{\text{local}}$, and the PL contrast at $\omega_{\text{local}}$ is required to extract the phase structure. Figure 1(b) shows a representative single frequency PL contrast image of plane-wave-like spin waves in the YIG films. Figure 1(c) and (d) characterize the LSMO film, with (c) showing a magnetic force microscopy image and (d) a single frequency PL contrast image of its spin waves. Both LSMO measurements clearly reveal the presence of antiferromagnetically coupled stripe domains. Figure 1(e-g) present a typical analysis of spin waves using full-frequency spectra. For each position, a selected microwave frequency range is scanned to obtain frequency-dependent data (supplementary Fig. S2 [38]). A fit of the spectra provides information on the location-dependent resonance frequency $\omega_{\text{local}}$ and PL at $\omega_{\text{local}}$. Figure 1(h) summarizes the measured spin wave as a function of the magnetic field from 10 to 193 G for the 100 nm YIG film. The spin wave dispersion relation and a match between the measurements and theoretical calculations are provided in the supplementary Fig. S3 [38].

Figure 2 depicts PL contrast images of spin waves in the YIG films interacting with magnetic point scatterers. While previous studies have visualized spin waves near film boundaries and single scatterers [48,55], we examine both single and multiple scatterers with varied incident wavelengths. Figure 2(a) shows the spin-wave near a single scatterer. When the spin wave periodicity, $2\pi/|\boldsymbol{k}|$, is comparable to the scatterer's linear dimension $r$, a prominent wavefront bending is observed. $r$ is directly measured from zoomed-in single frequency imaging. Conversely, when $2\pi/|\boldsymbol{k}| \gg r$, the spin wave passes by without a large distortion. Figure 2(b) plots the interaction near two scatterers, where a clear interference pattern emerges. As a magnetic point scatterer acts as a secondary radiation source and is equivalent to a microregion possessing $\boldsymbol{m}_{\text{scat}} = \boldsymbol{m}_0 e^{-\frac{\rho^2}{2r^2}} e^{i\boldsymbol{k}\cdot\boldsymbol{\rho}}$, when a coherent spin wave with dynamic magnetization component $\boldsymbol{m}$ passes through, the divergence equation including the impurity source is modified as [56,57]:

$$\nabla \cdot (\boldsymbol{h} + 4\pi\boldsymbol{m} + 4\pi\boldsymbol{m}_{\text{scat}}) = 0 \quad (4)$$

Calculating the relevant spatial Green's function, the total wave is

$$\psi_{\text{total}}(\boldsymbol{\rho}) = \psi_i(\boldsymbol{\rho}) + \psi_s(\boldsymbol{\rho})$$

$$\approx e^{i\boldsymbol{k}\cdot\boldsymbol{\rho}} + 2\pi\sqrt{2\pi}|\boldsymbol{m}_0|k^{\frac{1}{2}}r^2 \cos\theta \exp\left(-2k^2r^2\sin^2\left(\frac{\theta}{2}\right)\right)\frac{e^{i(\boldsymbol{k}\cdot\boldsymbol{\rho}+3\pi/4)}}{\sqrt{|\boldsymbol{\rho}|}} \quad (5)$$

where $\psi_i(\boldsymbol{\rho}) = e^{i\boldsymbol{k}\cdot\boldsymbol{\rho}}$ denotes the incident plane wave and $\theta$ is the scattering angle. The total phase shift near a magnetic defect depends on the amplitudes of the incident and scattered wave components. Detailed derivations are included in supplementary materials [38]. Figure 2(c) presents the calculations of scattered spin waves, where a phase modification is observed after the magnetic defect. Figure 2(d) illustrates the corresponding phase change. We note that the direction of phase accumulation is opposite to the propagation direction of the spin wave in real space, because spins near the source experience more oscillations. The scattering strength is governed by the ratio between the scatterer's size and the spin-wave wavelength, becoming significant only when these two length scales are comparable. The complex interferences provide a map of the relative phase and amplitude that can be used in designs of spin-wave devices [4].

To explore how extended magnetic structures, beyond point-like scatterers, influence spin waves, we studied their propagation in LSMO films with the same method used for YIG. Under low external magnetic fields, our LSMO films host antiferromagnetically coupled stripe domains under low external magnetic fields [27,28]. Previous studies have indicated that spin waves are most prominent when traveling along the spin stripe channels, since the propagation of spin waves becomes more difficult while passing boundaries between two domains with opposite magnetizations [28,31]. Figure 3(a) visualizes spin waves (vertically with a ~1 μm periodicity) traversing a background of magnetic stripes that tilt 5 degrees from the horizontal direction under an external magnetic field ranging from 271 to 290 G. Here we use NV's resonance away from the LSMO film as the reference to determine the extra local field. Figure 3(c) shows a clear wavefront modification that is dependent on the underlying domain structure. By isolating the signal at the local resonance frequency $\omega_{\mathrm{local}}$ to mitigate the background static domain contribution, we reveal a zig-zag-shaped distortion in the spin waves, which has not been previously reported. Figure 3(d) illustrates the spin-wave configurations within the stripe domains.

To investigate the origin of the altered spin waves, we performed micromagnetic simulations using Mumax3 [58]. We simulated a 160 nm-thick LSMO film with the following parameters: saturation magnetization $M_{\mathrm{S}} = 475\ \mathrm{kA/m}$, exchange stiffness $A_{\mathrm{ex}} = 1.94\ \mathrm{pJ/m}$, interfacial Dzyaloshinskii-Moriya interaction $D_{\mathrm{ind}} = 0.1\ \mathrm{mJ/m^2}$, $z$-axis uniaxial anisotropy constant $K_u = 27\ \mathrm{kJ/m^3}$, external magnetic field $B_{\mathrm{ext}} = 280\ \mathrm{G}$ along NV-axis; geometry is $8000 \times 432 \times 160\ \mathrm{nm^3}$; the mesh cell is $4 \times 4 \times 4\ \mathrm{nm^3}$; damping coefficient $\alpha$ is set to $5 \times 10^{-4}$ to obtain a clear and stable dispersion [28]. Antiferromagnetically coupled stripe domains are observed in Fig. 4(a). Both experiments [28] and micromagnetic simulations reveal that the magnetization within the stripe domains in LSMO points in the $\pm z$ direction and flips along the y-axis.

As plotted in Figs. 4(b) and 4(c), the dynamic magnetization components of the spin waves, $m_x$ and $m_y$, vary with position $(x,\ y)$. Both top and sectional views show that $m_x$ shares the same sign while $m_y$ has an opposite sign in the adjacent domains. The spin-wave dispersion is plotted in Fig. 4(d). In Fig. 4(e), we present a map of the spin-wave phase, calculated as $\arctan(m_y/m_x)$. Remarkably, a zig-zag wavefront is observed, which is an intrinsic physical property of spin-wave modes in the stripe domains, originating from the evolution of local $m_x$ and $m_y$. Figure 4(f) illustrates the spin-wave propagation in the up-and-down stripe domains. We find that the direction of the phase gradient is reversed in the two domains (Supplementary Materials Fig. S4). Following a procedure similar to that in Eqs. (2) and (3), we further estimate the NV signals by identifying a local minimum in $\Omega_R(\boldsymbol{\rho})$ while approximating $m_x$ and $m_y$ based on the simulated dynamic magnetizations. Results that are qualitatively consistent with the experiments are obtained and included in the Supplementary Materials Fig. S5 [38]. We note that the simulated dynamic magnetization and phase patterns are robust and do not require the application of a reference field, and we found that changing $M_{\mathrm{S}}$, $A_{\mathrm{ex}}$, $D_{\mathrm{ind}}$, $K_u$ by 20% independently has a negligible effect on the simulated zig-zag wavefronts.

The existence of antiferromagnetically coupled stripe domains has multiple effects on spin waves. As the equilibrium magnetization directions of adjacent stripes are opposite, spin waves must simultaneously satisfy both phase continuity and the dispersion relation near the domain boundaries, which forces the phase gradient directions in adjacent stripes to be different, producing a zig-zag wavefront. In combination with the modified wavefront, a ±7 G shift along the NV-axis is observed in the local field between neighboring stripes. This field variation affects the local NV resonance and is an experimental signal to identify the stripe polarity. We note that the stand-off distance from NV centers to the film surface in our experiment is estimated to be between 50 and 100 nm, which suggests an even more pronounced field shift on the film surface, possibly twice as large. At the stripe domain boundaries, the spins in the LSMO material undergo a continuous rotation from an out-of-plane to an in-plane orientation. This reorientation may give rise to localized boundary modes, which can act as scattering centers or waveguides, thereby influencing the coherence and propagation of the spin waves. In Figure 4, we find that $m_x$ is primarily concentrated near the domain boundaries for the spin-wave mode studied. Previous studies have suggested that the distinct spin environments in neighboring stripes may result in different spin-wave chiralities [28]. As the cross product of the local magnetic field and the spin magnetization changes sign in the LLG equation [38], the handedness of the spin waves changes. Our simulation shows that spin waves in adjacent domains with opposite polarities satisfy the sign change in the LLG equation. However, one may understand this behavior by a change in the phase propagation direction. Because the pure edge modes have a much higher frequency that is beyond the frequency range of the current work [28], and applying a larger external field to reach the desired frequency also changes the striped magnetic ground state in LSMO, future studies beyond the scope of the current work are necessary to fully clarify these points.

Our achievement of direct, real-space observation of spin-wave propagation near complex magnetic structures provides critical insights into spin-wave behaviors. This allows for the quantitative mapping of dispersion relations and the precise measurement of wave properties within actual devices. The propagation characteristics of spin waves are determined by the local spin structures through which they travel. Local barriers, such as magnetic inhomogeneities or domain walls, can scatter or guide spin waves, modifying their wavefront and coherence. NV spectroscopy establishes a direct method to track spin waves shaped by magnetic defects and domains [48,55]. Of course, NV imaging also has limitations. It may capture contributions from multi-mode spin excitations, and the measurement frequency is tied to the applied external field. Nonetheless, we envision that further developments to expand the measurement capability, e.g., off-resonance imaging to expand the frequency range [53,59,60] and quantum noise spectroscopy for capturing spin dynamics, can offer more comprehensive, spatially-resolved understanding of spin behaviors, assisting the design of spin wave logic elements with tailored functional responses [4,6,7].

Acknowledgments: We gratefully acknowledge technical assistance from C. Zhu, S. Li, and Z. Zhang at CIQTEK and Y. Wu at IOP. We thank K. Jin and H. Yuan at Zhejiang Univ. for helpful discussions and Y. Wu at Wuhan Univ. for help with the illustration figure of the experimental setup. This work was supported by the National Key Research and Development Program of China (No. 2022YFA1403400, No. 2023YFA1406500), the National Natural Science Foundation of China (No. 12274439, No. 52225205, No. 12404119, No. 12550004), Chinese Academy of Sciences through the Youth Innovation Promotion Association (No. 2022YSBR-048), Postdoctoral

Fellowship Program of CPSF (No. GZC20232944).

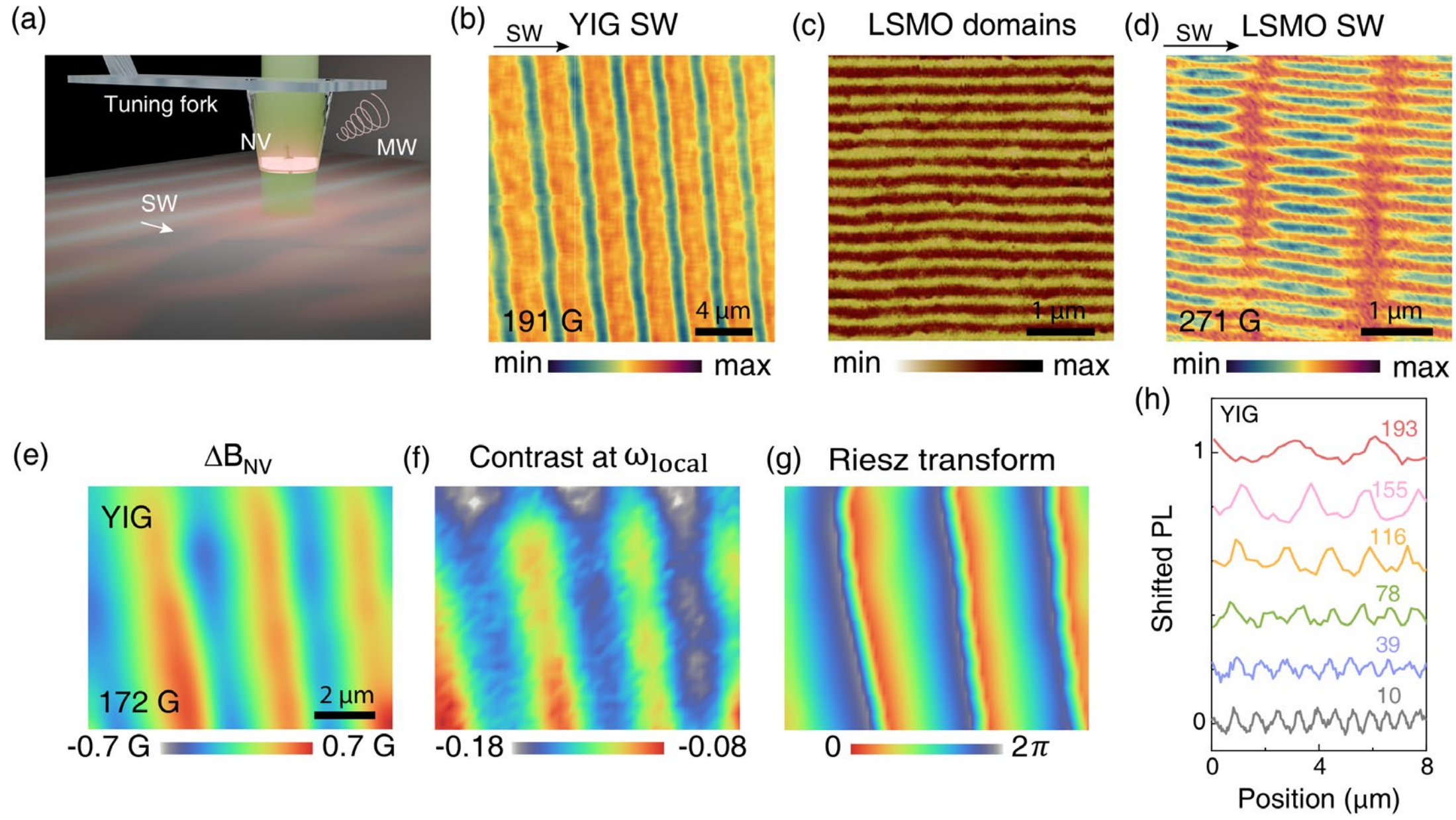

Fig. 1. Representative spin-wave and magnetic domain images of YIG and LSMO. (a) Schematic of the experimental setup. SW and MW label the spin-wave signal and microwave, respectively. (b) PL image of spin waves in the YIG film, acquired at a resonance frequency of external field, approximately 2335 MHz. (c) LSMO antiferromagnetically coupled stripe domains measured by a magnetic force microscope. (d) PL image of spin waves in the LSMO film, acquired at a resonance frequency of external field, approximately 2111 MHz. Both antiferromagnetically coupled stripe domains and spin waves are observed. (e-g) Images of magnetic field variations along the NV-axis, contrast at $\omega_{\mathrm{local}}$, and the Riesz transform associated with spin wavefront for spin waves in the YIG film, obtained by fitting the NV spectrum at each pixel. (h) Summary plot of one-dimensional cross-sections of spin waves in the YIG film at varied external magnetic fields. Black arrows labeled "SW" indicate the spin-wave propagation direction throughout this work.

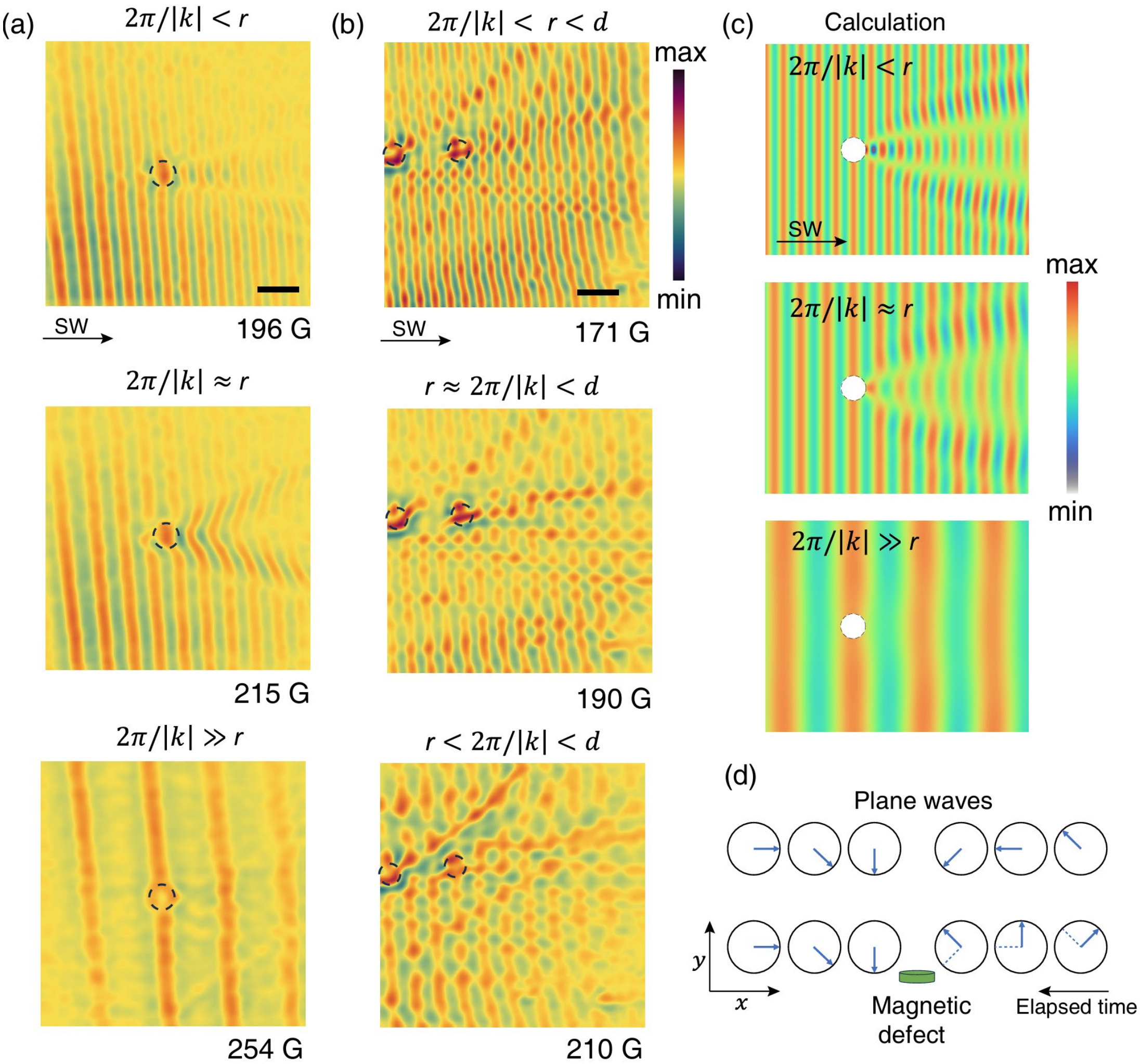

Fig. 2. Images of spin waves in the YIG film near magnetic point scatterers. (a) Spin waves with varied incident wavelengths near a single magnetic scatterer. Measured at frequencies of approximately 2321, 2267, 2158 MHz. (b) Spin waves with varied incident wavelengths near two magnetic scatterers. Measured at frequencies of approximately 2392, 2338, 2282 MHz. Dashed circles indicate the positions of the scatterers. Black bars indicate 5 $\mu$m. (c) Calculations for scattering of spin waves by a magnetic defect. (d) An illustration of the spin-wave phase modified by a magnetic defect. The total phase shift near a magnetic defect depends on the amplitudes of the incident and scattered wave components. The direction of phase accumulation is opposite to the propagation direction of the spin wave in real space, because spins near the source experience more oscillations.

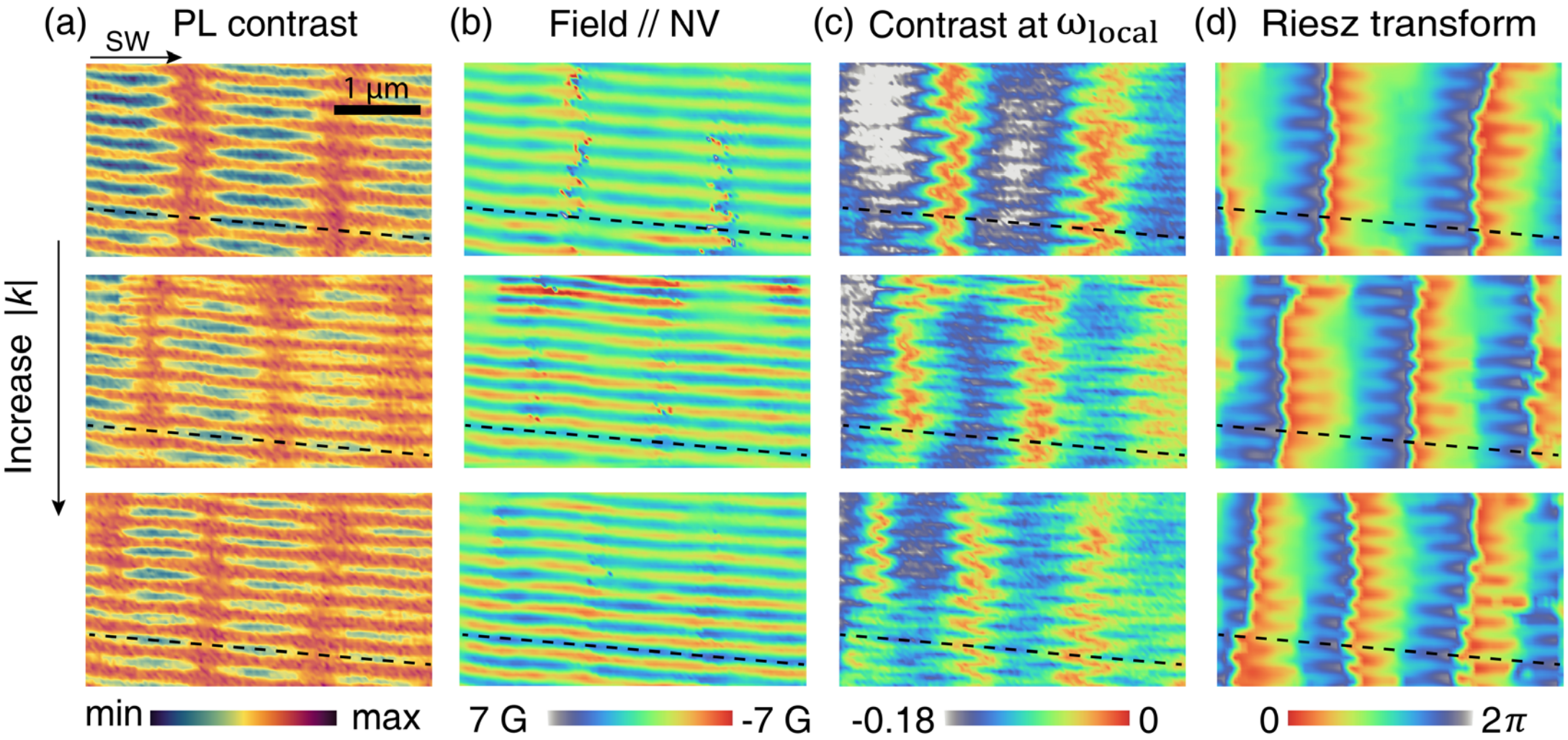

Fig. 3. NV measurements of antiferromagnetically coupled stripe domains and spin waves in LSMO. (a) PL contrast at the NV resonance frequency under different external magnetic fields, 271 G, 281 G, and 290 G (top to bottom panels). Measured at frequencies of approximately 2111, 2083, and 2057 MHz. (b) Corresponding maps of the magnetic field shift along the NV-axis caused by stripe domains. Measured by a frequency scan at each pixel. (c) Contrast at $\omega_{\mathrm{local}}$. (d) Results of the Riesz transform applied to the spin-wave wavefront. Black dashed lines are guides to the eye.

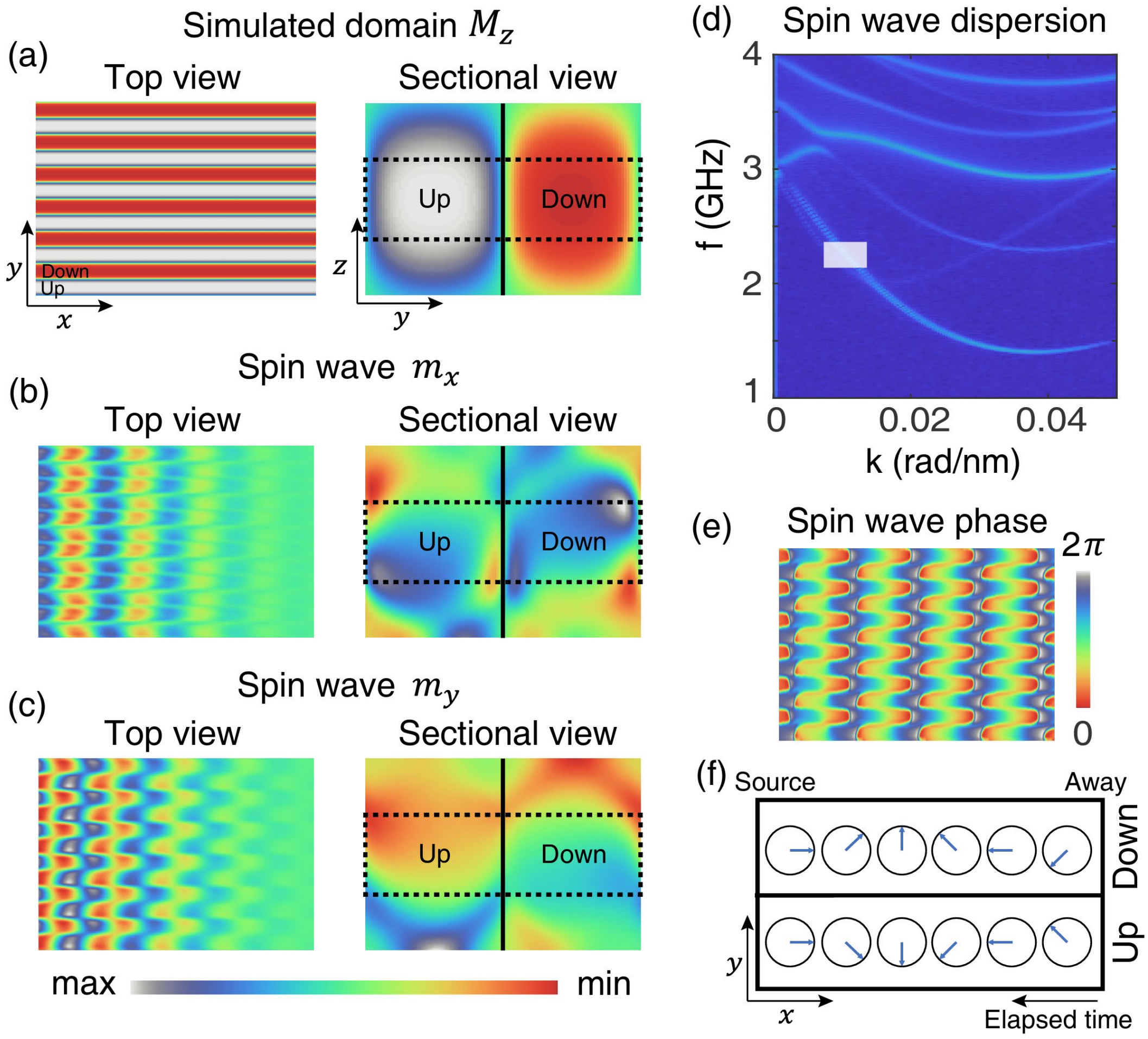

Fig. 4. Micromagnetic simulation and calculations. (a) Simulated ground state magnetization $M_Z$, showing the formation of stripe domains. (b, c) Simulated dynamic magnetization components $m_x$ and $m_y$ originating from spin-wave propagation at an arbitrary time, generated via the inverse Fourier transform of the frequency and wavevector selected by the white band in panel (d). Dashed areas indicate bulk-dominated behaviors. (d) Simulated spin-wave dispersion of LSMO, obtained by applying a double Fourier transform to a time series of the simulated dynamic magnetization. (e) Simulated spin-wave phase obtained from $m_x$ and $m_y$. (f) An illustration of spin-wave propagation in the adjacent domains, based on data in panels (a-c).

References:

[1] K. Wagner, A. Kákay, K. Schultheiss, A. Henschke, T. Sebastian, and H. Schultheiss, Magnetic domain walls as reconfigurable spin-wave nanochannels, Nat. Nanotechnol. **11**, 432 (2016).

[2] H. Yu, J. Xiao, and H. Schultheiss, Magnetic texture based magnonics, Phys. Rep. **905**, 1 (2021).

[3] H. Y. Yuan, Y. Cao, A. Kamra, R. A. Duine, and P. Yan, Quantum magnonics: when magnon spintronics meets quantum information science. Phys. Rep. **965**, 1-74 (2022).

[4] A. V. Chumak, V. I. Vasyuchka, A. A. Serga, and B. Hillebrands, Magnon spintronics, Nat. Phys. **11**, 453 (2015).

[5] D. Grundler, Nanomagnonics around the corner, Nat. Nanotechnol. **11**, 407 (2016).

[6] A. Khitun, M. Bao, and K. L. Wang, Magnonic logic circuits, J. Phys. D: Appl. Phys. **43**, 264005 (2010).

[7] A. Haldar, D. Kumar, and A. O. Adeyeye, A reconfigurable waveguide for energy-efficient transmission and local manipulation of information in a nanomagnetic device, Nat. Nanotechnol. **11**, 437 (2016).

[8] V. V. Kruglyak, S. O. Demokritov, and D. Grundler, Magnonics, J. Phys. D: Appl. Phys. **43**, 264001 (2010).

[9] A. A. Serga, A. V. Chumak, and B. Hillebrands, YIG magnonics, J. Phys. D: Appl. Phys. **43**, 264002 (2010).

[10] T. Jungwirth, X. Marti, P. Wadley, and J. Wunderlich, Antiferromagnetic spintronics, Nat. Nanotechnol. **11**, 231 (2016).

[11] V. Baltz, A. Manchon, M. Tsoi, T. Moriyama, T. Ono, and Y. Tserkovnyak, Antiferromagnetic spintronics, Rev. Mod. Phys. **90**, 015005 (2018).

[12] S.-H. Yang, R. Naaman, Y. Paltiel, and S. S. P. Parkin, Chiral spintronics, Nat. Rev. Phys. **3**, 328 (2021).

[13] A. Mudhafer, M. A. Najdi, R. M. Shaaban, N. J.Jassim, A. Ateeq, and M. T. Alshamkhani, Transformations of Stripy Magnetic Spin Texture: A Mini Review, Physica B: Condensed Matter 417886 (2025).

[14] Y. Zhou, S. Li, X. Liang, and Y. Zhou, Topological Spin Textures: Basic Physics and Devices, Adv. Mater. **37**, 2312935 (2025).

[15] L. Landau and E. Lifshitz, On the theory of the dispersion of magnetic permeability in ferromagnetic bodies, Phys. Z. Sowjetunion **8**, 101 (1935).

[16] T. L. Gilbert, Classics in Magnetics A Phenomenological Theory of Damping in Ferromagnetic Materials, IEEE Trans. Magn. **40**, 3443 (2004).

[17] J. Krčma, O. Wojewoda, M. Hrtoň, J. Holobrádek, J. A. Arregi, J. Panda, E. Pribytova, and M. Urbánek, Mie-enhanced microfocused brillouin light scattering for full wave vector resolution of nanoscale spin waves, Sci. Adv. **11,** eady8833 (2025).

[18] S. Wintz, V. Tiberkevich, M. Weigand, J. Raabe, J. Lindner, A. Erbe, A. Slavin, and J. Fassbender, Magnetic vortex cores as tunable spin-wave emitters, Nat. Nanotechnol. **11**, 948 (2016).

[19] V. Sluka, T. Schneider, R. A. Gallardo, A. Kákay, M. Weigand, T. Warnatz, R. Mattheis, A. Roldán-Molina, P. Landeros, V. Tiberkevich, A. Slavin, G. Schütz, A. Erbe, A. Deac, J. Lindner, J. Raabe, J. Fassbender, and S. Wintz, Emission and propagation of 1D and 2D spin waves with nanoscale wavelengths in anisotropic spin textures, Nat. Nanotechnol. **14**, 328 (2019).

[20] E. Albisetti, S. Tacchi, R. Silvani, G. Scaramuzzi, S. Finizio, S. Wintz, C. Rinaldi, M. Cantoni,

J. Raabe, G. Carlotti, R. Bertacco, E. Riedo, and D. Petti, Optically inspired nanomagnonics with nonreciprocal spin waves in synthetic antiferromagnets, Adv. Mater. **32**, 1906439 (2020).
[21] X.-X. Zhang, L. Li, D. Weber, J. Goldberger, K. F. Mak, and J. Shan, Gate-tunable spin waves in antiferromagnetic atomic bilayers, Nat. Mater. **19**, 838 (2020).
[22] J.-H. Park, E. Vescovo, H.-J. Kim, C. Kwon, R. Ramesh, and T. Venkatesan, Direct evidence for a half-metallic ferromagnet, Nature **392**, 794 (1998).
[23] M. Bowen, M. Bibes, A. Barthélémy, J.-P. Contour, A. Anane, Y. Lemaître, and A. Fert, Nearly total spin polarization in $La_{2/3}Sr_{1/3}MnO_3$ from tunneling experiments, Appl. Phys. Lett. **82**, 233 (2003).
[24] Y. Zhou, T. Guo, L. Han, L. Liao, W. He, C. Wan, C. Chen, Q. Wang, L. Qiao, H. Bai, W. Zhu, Y. Zhang, R. Chen, X. Han, F. Pan, and C. Song, Spin-torque–driven antiferromagnetic resonance, Sci. Adv. **10**, eadk7935 (2024).
[25] Y. Z. Wang, T. Y. Zhang, J. Dong, P. Chen, G. Q. Yu, C. H. Wan, and X. F. Han, Voltage-Controlled Magnon Transistor via Tuning Interfacial Exchange Coupling, Phys. Rev. Lett. **132**, 076701 (2024).
[26] J. Wang, H. Wang, J. Chen, W. Legrand, P. Chen, L. Sheng, J. Xia, G. Lan, Y. Zhang, R. Yuan, J. Dong, X. Han, J.-P. Ansermet, and H. Yu, Broad-wave-vector spin pumping of flat-band magnons, Phys. Rev. Applied **21**, 044024 (2024).
[27] Y. Zhang, J. Liu, Y. Dong, S. Wu, J. Zhang, J. Wang, J. Lu, A. Rückriegel, H. Wang, R. Duine, H. Yu, Z. Luo, K. Shen, and J. Zhang, Strain-Driven Dzyaloshinskii-Moriya Interaction for Room-Temperature Magnetic Skyrmions, Phys. Rev. Lett. **127**, 117204 (2021).
[28] Y. Zhang, L. Qiu, J. Chen, S. Wu, H. Wang, I. A. Malik, M. Cai, M. Wu, P. Gao, C. Hua, W. Yu, J. Xiao, Y. Jiang, H. Yu, K. Shen, and J. Zhang, Switchable long-distance propagation of chiral magnonic edge states, Nat. Mater. **24**, 69 (2025).
[29] Q. Qin, S. He, W. Song, P. Yang, Q. Wu, Y. P. Feng, and J. Chen, Ultra-low magnetic damping of perovskite La0.7Sr0.3MnO3 thin films, Appl. Phys. Lett. **110**, 112401 (2017).
[30] M. Cai, S. Wang, Y. Zhang, X. Bao, D. Shen, J. Ren, L. Qiu, H. Yu, Z. Luo, M. Kläui, S. Zhang, N. Jaouen, G. Van Der Laan, T. Hesjedal, K. Shen, and J. Zhang, Stabilization and Observation of Large-Area Ferromagnetic Bimeron Lattice, Phys. Rev. Lett. **135**, 116703 (2025).
[31] C. Liu, S. Wu, J. Zhang, J. Chen, J. Ding, J. Ma, Y. Zhang, Y. Sun, S. Tu, H. Wang, P. Liu, C. Li, Y. Jiang, P. Gao, D. Yu, J. Xiao, R. Duine, M. Wu, C.-W. Nan, J. Zhang, and H. Yu, Current-controlled propagation of spin waves in antiparallel, coupled domains, Nat. Nanotechnol. **14**, 691 (2019).
[32] A. Gruber, A. Dräbenstedt, C. Tietz, L. Fleury, J. Wrachtrup, and C. V. Borczyskowski, Scanning Confocal Optical Microscopy and Magnetic Resonance on Single Defect Centers, Science **276**, 2012 (1997).
[33] M. Guo, M. Wang, P. Wang, D. Wu, X. Ye, P. Yu, Y. Huang, F. Shi, Y. Wang, and J. Du, A flexible nitrogen-vacancy center probe for scanning magnetometry, Rev. Sci. Instrum. **92**, 055001 (2021).
[34] Y. Li, C. Liu, W. Cheng, and J. Li, Multiscale Approach for Unconventional Superconductors, Electromag. Sci. **2**, 1 (2024).
[35] L. Rondin, J.-P. Tetienne, T. Hingant, J.-F. Roch, P. Maletinsky, and V. Jacques, Magnetometry with nitrogen-vacancy defects in diamond, Rep. Prog. Phys. **77**, 056503 (2014).
[36] E. V. Levine, M. J. Turner, P. Kehayias, C. A. Hart, N. Langellier, R. Trubko, D. R. Glenn, R.

R. Fu, and R. L. Walsworth, Principles and techniques of the quantum diamond microscope, Nanophotonics **8**, 1945 (2019).

[37] J. Du, F. Shi, X. Kong, F. Jelezko, and J. Wrachtrup, Single-molecule scale magnetic resonance spectroscopy using quantum diamond sensors, Rev. Mod. Phys. **96**, 025001 (2024).

[38] See Supplemental Materials at [Url] for the experimental details, micromagnetic simulations and analysis, which Includes Refs. [39–45].

[39] B. A. Kalinikos and A. N. Slavin, Theory of dipole-exchange spin wave spectrum for ferromagnetic films with mixed exchange boundary conditions, J. Phys. C: Solid State Phys. **19**, 7013 (1986).

[40] M. Donahue, *Object-Oriented Micromagnetics Modeling Framework (OOMMF)*, https://doi.org/10.18434/T4/1502495.

[41] G. Venkat, D. Kumar, M. Franchin, O. Dmytriiev, M. Mruczkiewicz, H. Fangohr, A. Barman, M. Krawczyk, and A. Prabhakar, Proposal for a standard micromagnetic problem: Spin wave dispersion in a magnonic waveguide, IEEE Trans. Magn. **49**, 524 (2013).

[42] Z. Ren, L. Jin, T. Wen, Y. Liao, X. Tang, H. Zhang, and Z. Zhong, MuFA (multi-type fourier analyzer): A tool for batch generation of MuMax3 input scripts and multi-type fourier analysis from micromagnetic simulation output data, Comput. Phys. Commun. **244**, 311 (2019).

[43] M. Beg, M. Lang, and H. Fangohr, Ubermag: Toward more effective micromagnetic workflows, IEEE Trans. Magn. **58**, 1 (2022).

[44] P. Gruszecki, M. Kasprzak, A. E. Serebryannikov, M. Krawczyk, and W. Śmigaj, Microwave excitation of spin wave beams in thin ferromagnetic films, Sci. Rep. **6**, 22367 (2016).

[45] Y. Zhang, T. Yu, J. Chen, Y. Zhang, J. Feng, S. Tu, and H. Yu, Antenna design for propagating spin wave spectroscopy in ferromagnetic thin films, J. Magn. Magn. Mater. **450**, 24 (2018).

[46] M. Borst, P. H. Vree, A. Lowther, A. Teepe, S. Kurdi, I. Bertelli, B. G. Simon, Y. M. Blanter, and T. van der Sar, Observation and control of hybrid spin-wave–Meissner-current transport modes, Science **382**, 430 (2023).

[47] I. Bertelli, J. J. Carmiggelt, T. Yu, B. G. Simon, C. C. Pothoven, G. E. W. Bauer, Y. M. Blanter, J. Aarts, and T. Van Der Sar, Magnetic resonance imaging of spin-wave transport and interference in a magnetic insulator, Sci. Adv. **6**, eabd3556 (2020).

[48] T. X. Zhou, J. J. Carmiggelt, L. M. Gächter, I. Esterlis, D. Sels, R. J. Stöhr, C. Du, D. Fernandez, J. F. Rodriguez-Nieva, F. Büttner, E. Demler, and A. Yacoby, A magnon scattering platform, Proc. Natl. Acad. Sci. U.S.A. **118**, e2019473118 (2021).

[49] B. G. Simon, S. Kurdi, J. J. Carmiggelt, M. Borst, A. J. Katan, and T. Van Der Sar, Filtering and Imaging of Frequency-Degenerate Spin Waves Using Nanopositioning of a Single-Spin Sensor, Nano Lett. **22**, 9198 (2022).

[50] I. Bertelli, B. G. Simon, T. Yu, J. Aarts, G. E. W. Bauer, Y. M. Blanter, and T. Van Der Sar, Imaging Spin‐Wave Damping Underneath Metals Using Electron Spins in Diamond, Adv. Quantum Tech. **4**, 2100094 (2021).

[51] R. Timalsina, H. Wang, B. Giri, A. Erickson, X. Xu, and A. Laraoui, Mapping of Spin‐Wave Transport in Thulium Iron Garnet Thin Films Using Diamond Quantum Microscopy, Adv. Elect. Materials **10**, 2300648 (2024).

[52] C. Lüthi, L. Colombo, F. Vilsmeier, and C. Back, Long-range spin wave imaging with nitrogen vacancy centers and time resolved magneto-optical measurements, Review of Scientific Instruments **96**, 033703 (2025).

[53] K. Ogawa, M. Tsukamoto, Y. Mori, D. Takafuji, J. Shiogai, K. Ueda, J. Matsuno, K. Sasaki, and K. Kobayashi, Wideband wide-field imaging of spin-wave propagation using diamond quantum sensors, Phys. Rev. Applied **23**, 054001 (2025).

[54] A. Dréau, M. Lesik, L. Rondin, P. Spinicelli, O. Arcizet, J.-F. Roch, and V. Jacques, Avoiding power broadening in optically detected magnetic resonance of single NV defects for enhanced dc magnetic field sensitivity, Phys. Rev. B **84**, 195204 (2011).

[55] H. Wang, M. Madami, J. Chen, H. Jia, Y. Zhang, R. Yuan, Y. Wang, W. He, L. Sheng, Y. Zhang, J. Wang, S. Liu, K. Shen, G. Yu, X. Han, D. Yu, J.-P. Ansermet, G. Gubbiotti, and H. Yu, Observation of Spin-Wave Moiré Edge and Cavity Modes in Twisted Magnetic Lattices, Phys. Rev. X **13**, 021016 (2023).

[56] J. Stigloher, M. Decker, H. S. Körner, K. Tanabe, T. Moriyama, T. Taniguchi, H. Hata, M. Madami, G. Gubbiotti, K. Kobayashi, T. Ono, and C. H. Back, Snell's law for spin waves, Phys. Rev. Lett. **117**, 037204 (2016).

[57] N. Loayza, M. B. Jungfleisch, A. Hoffmann, M. Bailleul, and V. Vlaminck, Fresnel diffraction of spin waves, Phys. Rev. B **98**, 144430 (2018).

[58] A. Vansteenkiste, J. Leliaert, M. Dvornik, M. Helsen, F. Garcia-Sanchez, and B. Van Waeyenberge, The design and verification of MuMax3, AIP Advances **4**, 107133 (2014).

[59] R. Li, C.-J. Wang, Z. Cheng, P. Wang, Y. Wang, C. Duan, H. Liu, F. Shi, and J. Du, Wideband microwave magnetometry using a nitrogen-vacancy center in diamond, Phys. Rev. A **99**, 062328 (2019).

[60] K. Ogawa, S. Nishimura, K. Sasaki, and K. Kobayashi, Demonstration of highly sensitive wideband microwave sensing using ensemble nitrogen-vacancy centers, Appl. Phys. Lett. **123**, 214002 (2023).